\documentclass[12pt]{iopart}
\usepackage{graphicx}
\newcommand{\eps}{\varepsilon}
\newcommand{\vphi}{\varphi}
\newcommand{\dis}{\displaystyle}
\begin{document}

\title[Analytic calculation of nonadiabatic transition probabilities]{
Analytic calculation of nonadiabatic transition probabilities 
from monodromy of differential equations}

\author{T Kato\dag\, K Nakamura\dag and M Lakshmanan\ddag}

\address{\dag\ Department of Applied Physics, 
Osaka City University, Sumiyoshi-ku, Osaka 558-8585, Japan}

\address{\ddag\ Centre for Nonlinear Dynamics and Department of Physics,
Bharathidasan University, Tiruchirapalli - 620 024, India}

\eads{\mailto{kato@a-phys.eng.osaka-cu.ac.jp},
\mailto{nakamura@a-phys.eng.osaka-cu.ac.jp} and
\mailto{lakshman25@satyam.net.in}}

\begin{abstract}

The nonadiabatic transition probabilities in the two-level systems
are calculated analytically by using the monodromy matrix determining
the global feature of the underlying differential equation. We study the
time-dependent $2 \times 2$ Hamiltonian with the tanh-type plus 
sech-type energy difference and with constant off-diagonal elements
as an example to show the efficiency of the monodromy approach.
The application of this method to multi-level systems is also
discussed.

\end{abstract}




\section{Introduction}
\label{sec:intro}

Analytic calculation of the time evolution in two-level systems 
has been studied by a number of authors for a long time
since the beginning years of quantum 
mechanics~\cite{Landau32,Zener32,Rosen32,Rabi37,Nikitin62,Demkov63,Demkov69,
Bambini81,Bambini84,Hioe84,Hioe85a,Hioe85b,Carroll86,Ishkhanyan00}. 
These results have been applied
to various areas of physics including quantum optics,
laser spectroscopy, nuclear magnetic resonance and atomic 
collisions~\cite{Allen75,Nikitin84,Oppo96,Nakamura02}. 
The importance of the study of quantum time-evolution
is still increasing even now; For example, much attention has been 
paid recently to the quantum manipulation of qubits~\cite{Nielsen00}
and magnetization process of magnetic molecules with 
large spin~\cite{Chudnovsky98}.
Recent rapid development of computers has enabled massive
numerical simulation of quantum dynamics. Nevertheless, 
it remains to be important to study analytically solvable 
models for the following reasons: (1) in some ranges of physical 
parameters the numerical simulation becomes too difficult, and (2) analytic
solutions give a clearer description about parameter dependence.

Analytic solutions of quantum dynamics can be classified into 
several classes. Some of them are obtained by using 
hypergeometric functions. This was first found by
Rosen and Zener~\cite{Rosen32}, which has then been generalized by
several authors~\cite{Demkov69,Bambini81,Bambini84,Hioe84,Hioe85a,Hioe85b}. 
In these studies, the time-variable $t$ is generally
transformed into another 
real variable $z=z(t)$, which varies from $0$ 
to $1$ monotonically. 
Then, the Schr\"odinger equation of two-level systems
can be reduced to the hypergeometric differential equation, and
the transition probability can be related to the connection
problem between two pairs of fundamental solutions around
$z=0$ and $z=1$. 

One exception is the approach by Carroll and
Hioe~\cite{Carroll86}. They have studied two solvable classes,
and in one of them they have introduced
a new variable $z(t)$ changing from $-\infty$ to $\infty$ 
as $t$ increases and have reduced the Schr\"odinger equation 
to the Riemann-Papperitz equation. Recently, Ishkhanyan has pointed
out that the Carroll-Hioe model can be understood 
in terms of the hypergeometric functions by considering
a complex-valued path $z(t)=(y(t)+{\rm i})/2{\rm i}$ where
$y(t)$ is a real variable~\cite{Ishkhanyan00}. 
By this complex-valued path, Ishkhanyan also found 
a new solvable class, but he did not obtain results 
for the transition probability.

In this paper, we show that for the complex-valued path, 
the transition probability
can be calculated efficiently from the `monodromy' matrices
of the corresponding differential equations.
Monodromy is one of the global
properties of differential equations, and
has attracted much attention by mathematicians, for example,
through the deep connection with the Painlev\'e equations~\cite{Iwasaki91}.
Hence, we expect that the monodromy approach is valuable
not only because it enables
one to calculate the transition probability for various models
but also because it establishes a connection between
physical phenomena and global features of differential equations.

In this paper, as a concrete example, we mainly consider the following 
time-dependent two-level Schr\"odinger equation and obtain the
transition probability using the monodromy associated with the solution:
\begin{equation}
{\rm i} \left( \begin{array}{c} a_{1t} \\ a_{2t} \end{array} \right)
= \left( \begin{array}{cc} \eps(t) & V(t) \\ V(t) & -\eps(t) \end{array}
\right)
\left( \begin{array}{c} a_1 \\ a_2 \end{array} \right),
\label{eq:Sch}
\end{equation}
where the matrix elements are given by
\begin{eqnarray}
\eps(t) &=& E_0 \ {\rm sech} (t/T) + E_1 \tanh(t/T), 
\label{eq:eps} \\
V(t) &=& V_0.
\label{eq:v}
\end{eqnarray}
Here, the coefficients, $E_0$, $E_1$ and $V_0$, are assumed to
be real constants. This is one of the solvable class reported
by Ishkhanyan~\cite{Ishkhanyan00}. 
However, the transition probability for the model
has not been obtained. It should be
noted that this model is equivalent to the Rosen-Zener 
model~\cite{Rosen32}
in the case $E_1=0$
\footnote{In the original Rosen-Zener model, $\eps(t)$
is a constant, while $V(t)$ has a sech-type pulse form. This Hamiltonian,
however, is reformed by a proper unitary transformation of the wave function
to coincide with the present model.}, 
and that it also includes the special case
of the second Demkov-Kunike model~\cite{Demkov69} in the case $E_0=0$.
Hence, this model can give a smooth connection between the two known
results. 

The plan of the paper is as follows.
We give the relation between the transition probability
and the monodromy of the hypergeometric function
in \S~\ref{sec:monodromy}, and the transition probability
is calculated explicitly in \S~\ref{sec:calculation}. 
The extension to the multi-level problems
is addressed in \S~\ref{sec:discussion}. Finally, the results are 
summarized in \S~\ref{sec:summary}. In \ref{app:class},
we describe the generalization of the present model
and its relationship to the Carroll-Hioe's model.

\section{Hypergeometric function and monodromy}
\label{sec:monodromy}

The diagonal elements in the model (\ref{eq:Sch}) 
are eliminated by the following change of variables:
\begin{eqnarray}
c_1 &=& a_1 \exp \left( {\rm i}\int_0^t \eps {\rm d}t \right), \\
c_2 &=& a_2 \exp \left( -{\rm i}\int_0^t \eps {\rm d}t \right).
\end{eqnarray}
Then, the Schr\"odinger equation is expressed as
\begin{eqnarray}
{\rm i} c_{1t} &=& V
\exp \left( 2{\rm i}\int_0^t \eps {\rm d}t \right) c_2, \\
{\rm i} c_{2t} &=& V
\exp \left( -2{\rm i}\int_0^t \eps {\rm d}t \right) c_1.
\end{eqnarray}
By combining these two equations, we obtain the second-order
differential equations for $c_1$ and $c_2$ respectively as
\begin{eqnarray}
& & c_{1tt} + \left( -2{\rm i}\eps(t) - \frac{V_t}{V} \right) c_{1t}
+ V^2 c_1 = 0,
\label{eq:2nd-c1} \\
& & c_{2tt} + \left( 2{\rm i}\eps(t) - \frac{V_t}{V} \right) c_{2t}
+ V^2 c_2 = 0.
\label{eq:2nd-c2}
\end{eqnarray}
It should be noted that the equation for $c_2$ is obtained by replacing
$\eps(t)$ by $-\eps(t)$ in (\ref{eq:2nd-c1}). Hence, once the solution
of the equation for $c_1$ is obtained, the solution for $c_2$ is easily
obtained by reversing the sign of the parameters in $\eps(t)$.

The above discussion is general. Now, we consider the specific model
given by (\ref{eq:eps}) and (\ref{eq:v}). By substituting these
specific forms of $\eps(t)$ and $V(t)$
into (\ref{eq:2nd-c1}) and adopting the change of variable as
\begin{equation}
z(t) = \frac{\sinh(t/T) + {\rm i}}{2{\rm i}},
\label{eq:zt}
\end{equation}
equation (\ref{eq:2nd-c1}) can be reduced to the
differential equation of the hypergeometric function~\cite{Murphy60},
\begin{equation}
z(1-z) c_{1zz} + (\gamma - (1 + \alpha + \beta)z ) c_{1z} - \alpha\beta
c_1 = 0.
\label{eq:Gauss}
\end{equation}
Here, the parameters, $\alpha$, $\beta$ and $\gamma$,
are determined as
\begin{eqnarray}
\alpha &=& {\rm i}T(-E_1 + \sqrt{E_1^2 + V_0^2} ), 
\label{eq:alpha} \\
\beta  &=& {\rm i}T(-E_1 - \sqrt{E_1^2 + V_0^2} ),
\label{eq:beta} \\
\gamma &=& \frac12 + E_0 T - {\rm i}E_1 T.
\label{eq:gamma}
\end{eqnarray}
In the same way, equation (\ref{eq:2nd-c2}) is reduced to
the hypergeometric differential equation with the parameters
\begin{eqnarray}
\alpha' &=& {\rm i}T( E_1 + \sqrt{E_1^2 + V_0^2} ), \\
\beta'  &=& {\rm i}T( E_1 - \sqrt{E_1^2 + V_0^2} ), \\
\gamma' &=& \frac12 - E_0 T + {\rm i}E_1 T.
\end{eqnarray}
As already mentioned, these parameters are obtained by
replacing $E_0$ and $E_1$ by $-E_0$ and $-E_1$ respectively
in (\ref{eq:alpha})-(\ref{eq:gamma}). In the following calculation,
related to the calculation for $c_2$, the prime indicates that 
it is obtained by reversing the sign of $E_0$ and $E_1$ from
the original quantity without the prime.

From (\ref{eq:zt}), it can be easily seen that
the variable $|z| \rightarrow \infty$ as $|t| \rightarrow \infty$.
Hence, for discussion about the initial state it is convenient
to consider the fundamental solutions around $z=\infty$ as
\begin{eqnarray}
c_1 &=& A_1 f_{\infty}(z;\alpha) + A_2 f_{\infty}(z;\beta), 
\label{eq:c1-inf}
\\
c_2 &=& B_1 f_{\infty}(z;\alpha') + B_2 f_{\infty}(z;\beta'), 
\label{eq:c2-inf}
\end{eqnarray}
where $f_{\infty}(z;\alpha)$ and $f_{\infty}(z;\beta)$ are expressed
in terms of the hypergeometric functions as
\begin{eqnarray}
f_{\infty}(z;\alpha) &=& z^{-\alpha} F(\alpha,\alpha-\gamma+1,
\alpha-\beta+1;1/z), \\
f_{\infty}(z;\beta) &=& z^{-\beta} F(\beta-\gamma+1,\beta,
\beta-\alpha+1;1/z),
\end{eqnarray}
with similar definitions for $f_{\infty}(z;\alpha')$ and 
$f_{\infty}(z,\beta')$.
Since $\alpha$ and $\beta$ are pure-imaginary in the present model, 
we have to choose $\arg(z)$ to determine
the branch. In this paper, we choose
\begin{equation}
\arg(z) = \left\{ \begin{array}{ll}
\pi/2 & (t \rightarrow -\infty) \\
-\pi/2 & (t \rightarrow +\infty)
\end{array}. \right.
\end{equation}
In order to decide the initial state, it is sufficient to study
the limit $|z| \rightarrow \infty$, in which case we obtain
\begin{eqnarray}
c_1 &\rightarrow& A_1 z^{-\alpha} + A_2 z^{-\beta},
\label{eq:c1-inf2} \\
c_2 &\rightarrow& B_1 z^{-\alpha'} + B_2 z^{-\beta'}.
\label{eq:c2-inf2}
\end{eqnarray}
From (\ref{eq:zt}), (\ref{eq:c1-inf2}) and (\ref{eq:c2-inf2}), 
$A_i$'s and $B_i$'s are determined.

To obtain the transition probability, we assume that
the initial state is the ground state of the Hamiltonian 
in the limit $t \rightarrow -\infty$,
\begin{equation}
H = \left( \begin{array}{cc} -E_1 & V_0 \\
V_0 & E_1 \end{array} \right).
\label{eq:Ham--inf}
\end{equation}
It may be noted
that the off-diagonal elements in (\ref{eq:Ham--inf}) 
do not vanish. Consequently the
ground state wavefunction does not correspond to $|a_1| = 1$ 
and $a_2 = 0$ as it appears in the usual models.
The time-evolution of the ground-state wave function is obtained
generally as
\begin{equation}
\left( \begin{array}{c} a_1(t) \\ a_2(t) \end{array} \right)
= \left( \begin{array}{c} A \\ -A' \end{array} \right)
e^{-{\rm i} \left( - \sqrt{E_1^2 + V_0^2} \right)t + {\rm i}\vphi},
\label{eq:a-evinf}
\end{equation}
where
\begin{equation}
A = \sqrt{\frac{E_1 + \sqrt{E_1^2 + V_0^2}}{2\sqrt{E_1^2+V_0^2}}}, 
\hspace{5mm}
A' = \sqrt{\frac{-E_1 + \sqrt{E_1^2 + V_0^2}}{2\sqrt{E_1^2+V_0^2}}}. 
\label{eq:A-Aprime}
\end{equation}
The solution (\ref{eq:a-evinf}) includes an arbitary phase factor $\vphi$, 
which is chosen zero in this paper. From (\ref{eq:a-evinf}), 
the time-evolutions
of $c_1$ and $c_2$ in the limit $t\rightarrow -\infty$ can be 
easily evaluated as
\begin{eqnarray}
c_1(t) &=& a_1(t) \exp\left({\rm i}\int_0^t \eps {\rm d}t\right)
\rightarrow A e^{-{\rm i}
\left(E_1 - \sqrt{E_1^2 + V_0^2}\right)t-{\rm i}\phi_0 - 
{\rm i}\phi_1}, 
\label{eq:c1t-inf} \\
c_2(t) &=& a_2(t) \exp\left(-{\rm i}\int_0^t \eps {\rm d}t\right)
\rightarrow -A' e^{-{\rm i}
\left(-E_1 - \sqrt{E_1^2 + V_0^2}\right)t+{\rm i}\phi_0 +
{\rm i}\phi_1},
\label{eq:c2t-inf}
\end{eqnarray}
where the phase factors are given as
\begin{equation}
\phi_0 = TE_0 \frac{\pi}{2}, \hspace{5mm}
\phi_1 = TE_1 \log 2.
\end{equation}
By using (\ref{eq:zt}) and by comparing 
(\ref{eq:c1t-inf})-(\ref{eq:c2t-inf}) with 
(\ref{eq:c1-inf2})-(\ref{eq:c2-inf2}),
the constants, $A_i$'s and $B_i$'s are determined as
\begin{eqnarray}
A_1 = A e^{{\rm i}\pi \alpha/2 - {\rm i}\phi_1 - {\rm i}\phi_0
- {\rm i} \vphi_1}, & \hspace{5mm}& A_2 = 0, 
\label{eq:A1} \\
B_1 = -A' e^{{\rm i}\pi \alpha'/2 + {\rm i}\phi_1 + {\rm i}\phi_0
- {\rm i} \vphi'_1}, &\hspace{5mm} & B_2 = 0,
\label{eq:B1}
\end{eqnarray}
where the phase factors, $\vphi_1$ and $\vphi'_1$, are given as
\begin{equation}
{\rm i}\vphi_1 = 2 \alpha \log 2, \hspace{5mm}
{\rm i}\vphi'_1 = 2 \alpha' \log 2,
\end{equation}
though these are not relevant to the calculation of the
transition probability.

\begin{figure}
\begin{center}
\includegraphics[width=75mm]{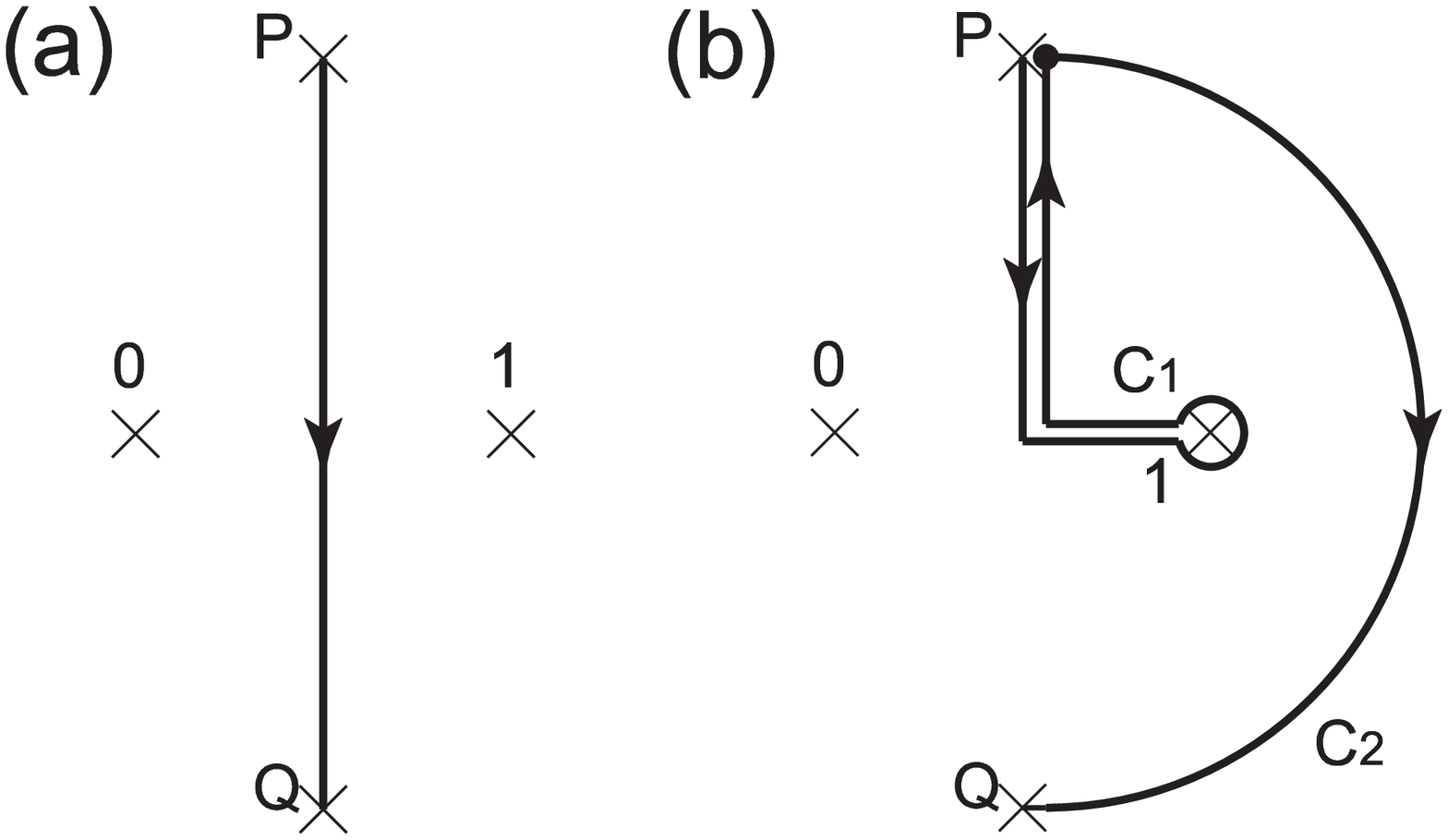}
\end{center}
\caption{\label{fig:path} The path along which the analytic continuation
is performed; (a) The original path and (b) deformed path.}
\end{figure}

By the choice of the initial condition, the time-evolution has 
been described only
by the fundamental solution $f_{\rm \infty}(z;\alpha)$ around
$z = \infty$. To be more accurate,
around $z= {\rm i}\infty + 1/2$ (corresponding to 
$t \rightarrow -\infty$) denoted by the point P in
Fig.~\ref{fig:path}~(a), the solution is given by
\begin{eqnarray}
c_1(z) &=& A_1 f_{\infty}(z;\alpha), 
\label{eq:pointPc1} \\
c_2(z) &=& B_1 f_{\infty}(z;\alpha').
\label{eq:pointPc2}
\end{eqnarray}
On the other hand, the final state is given by the solution
of the hypergeometric differential equation 
around $z =-{\rm i}\infty+1/2$ (corresponding to 
$t\rightarrow \infty$) denoted by the point Q
in Fig.~\ref{fig:path}~(a). The path $z(t)$ in the complex plane
is also drawn in Fig.~\ref{fig:path}.
Then, the solution around the point Q analytically continued from 
the point P does not equal to (\ref{eq:pointPc1}) and (\ref{eq:pointPc2});
The solution is expressed as linear combinations of the 
fundamental solutions around $z=\infty$. This is crucial to the calculation
of the transition probability. 

In order to make the situation clearer, let us deform the path of $z$
as shown in Fig.~\ref{fig:path}~(b). In this deformed path, 
the analytic continuation of the solution is divided into two parts, 
${\rm C}_1$ and ${\rm C}_2$. Here, the path ${\rm C}_1$ denotes a 
round trip to the singular point at $z = 1$, while the path ${\rm C}_2$
is a half round trip around $z=\infty$ in the clockwise direction.
The analytic continuation along the path ${\rm C}_2$ is easily
performed, and determined only by the fundamental solutions
around $z=\infty$. On the other hand, the analytic continuation along
the path ${\rm C}_1$ is nontrivial, and determined by the global
character of the differential equation called the `monodromy'. 
The monodromy is expressed by the monodromy matrices as
\begin{eqnarray}
\gamma({\rm C}_1) \left(f_{\infty}(z;\alpha), f_{\infty}(z;\beta)\right)
&=& \left(f_{\infty}(z;\alpha), f_{\infty}(z;\beta)\right) R, 
\label{eq:orig-mono} \\
\gamma({\rm C}_1) \left(f_{\infty}(z;\alpha), f_{\infty}(z;\beta)\right)
&=& \left(f_{\infty}(z;\alpha'), f_{\infty}(z;\beta')\right) R',
\end{eqnarray}
where $\gamma({\rm C}_1)$ denotes the analytic continuation along
the path ${\rm C}_1$. Denoting the matrix elements of $Q$ and $Q'$ as
\begin{equation}
R = \left( \begin{array}{cc} a & b \\ c & d \end{array} \right),
\hspace{5mm}
R' = \left( \begin{array}{cc} a' & b' \\ c' & d' \end{array} \right),
\label{eq:orig-ele}
\end{equation}
the solutions around the point Q can be expressed by
\begin{eqnarray}
\gamma({\rm C}_1) \left( c_1(z) \right)
&=& \gamma({\rm C}_1) \left( A_1 f_{\infty}(z;\alpha) \right)
\nonumber \\
&=& A_1 a f_{\infty}(z;\alpha) + A_1 c f_{\infty}(z;\beta), 
\label{eq:c1-connect} \\
\gamma({\rm C}_1) \left( c_2(z) \right)
&=& \gamma({\rm C}_1) \left( B_1 f_{\infty}(z;\alpha') \right)
\nonumber \\
&=& B_1 a' f_{\infty}(z;\alpha') + B_1 c' f_{\infty}(z;\beta').
\label{eq:c2-connect}
\end{eqnarray}
Here, as shown below, the first (second) term in the final entries of
(\ref{eq:c1-connect})
and (\ref{eq:c2-connect}) corresponds to the excited (ground)
state in the limit $t \rightarrow \infty$. Hence, in the calculation
of the transition probability, only the element $a$ ($a'$) is relevant. 
This matrix element is calculated explicitly in the next section.

Let us end this section by deriving the formula for the 
transition probability, using the monodromy matrix elements. In the limit 
$t \rightarrow \infty$, the equations, (\ref{eq:c1-connect})
and (\ref{eq:c2-connect}), are evaluated as
\begin{eqnarray}
c_1(t) &\rightarrow& a A_1 e^{{\rm i}\vphi + {\rm i}\pi \alpha/2}
e^{-{\rm i}\left(E_1+\sqrt{E_1^2 + V_0^2}\right)t} + \cdots, \\
c_2(t) &\rightarrow& a B_1 e^{{\rm i}\vphi' + {\rm i}\pi \alpha'/2}
e^{-{\rm i}\left(-E_1+\sqrt{E_1^2 + V_0^2}\right)t} + \cdots.
\end{eqnarray}
Here, we have suppressed the second term 
corresponding to the ground state. 
From these equations, the components of the wave function 
$\Psi(t) = (a_1(t), a_2(t))^{{\rm T}}$ is obtained in the limit
$t\rightarrow \infty$ as
\begin{eqnarray}
a_1(t) &=& c_1(t) \exp \left\{ -{\rm i}\int_0^t \eps(t) {\rm d}t \right\}
\nonumber \\
&\rightarrow& a A_1 e^{{\rm i}\vphi + {\rm i}\pi \alpha + {\rm i}\phi_1
- {\rm i}\phi_0} e^{-{\rm i}\sqrt{E_1^2 + V_0^2}t} + \cdots , \\
a_2(t) &=& c_2(t) \exp \left\{ +{\rm i}\int_0^t \eps(t) {\rm d}t \right\}
\nonumber \\
&\rightarrow& a' B_1 e^{{\rm i}\vphi' + {\rm i}\pi \alpha' - {\rm i}\phi_1
+ {\rm i}\phi_0} e^{-{\rm i}\sqrt{E_1^2 + V_0^2}t} + \cdots .
\end{eqnarray}
By substituting (\ref{eq:A1}) and (\ref{eq:B1}), 
the wave function $\Psi(t)$ is evaluated as
\begin{equation}
\Psi(t) \rightarrow \left( \begin{array}{c}
a A e^{-2{\rm i}\phi_0 + {\rm i}\pi \alpha} \\
- a' A' e^{2{\rm i}\phi_0 + {\rm i}\pi \alpha'}
\end{array} \right) e^{-{\rm i}\sqrt{E_1^2+V_0^2}t} + \cdots .
\end{equation}
On the other hand, the Hamiltonian and the wave function of 
the excited state $\Psi_{\rm E.S.}$ in the 
limit $t \rightarrow \infty$ are given as
\begin{equation}
H = \left( \begin{array}{cc} E_1 & V_0 \\
V_0 & -E_1 \end{array} \right), \hspace{5mm} 
\Psi_{{\rm E. S.}} = \left( \begin{array}{c} A \\ A' 
\end{array} \right),
\end{equation}
where $A$ and $A'$ are given by (\ref{eq:A-Aprime}).
Then, the transition probability is calculated as
\begin{eqnarray}
P &=& \left| \Psi(t\rightarrow \infty) 
\cdot \Psi_{{\rm E. S.}} \right|^2 \nonumber \\
&=& \left| a A^2 e^{{\rm i}\pi \alpha -2{\rm i}\phi_0}
- a' A'^2 e^{{\rm i}\pi \alpha' + 2{\rm i}\phi_0} \right|^2 \nonumber \\
&=& \Bigg| \frac{\eps_1}{2\sqrt{\eps_1^2+v^2}}
\left( a e^{\eps_1 - \sqrt{\eps_1^2+v^2}} e^{-{\rm i}\eps_0}
+ a'e^{-\eps_1-\sqrt{\eps_1^2+v^2}} e^{{\rm i}\eps_0} \right) \Biggr. 
\nonumber \\
& & \Biggl. + \frac12 \left( a e^{\eps_1 - \sqrt{\eps_1^2 + v^2}}
e^{-{\rm i}\eps_0} - a' e^{-\eps_1 - \sqrt{\eps_1^2 + v^2}} e^{{\rm i}
\eps_0} \right) \Biggr|^2,
\label{eq:formula}
\end{eqnarray}
where in the final equation we have introduced new variables,
\begin{equation}
\eps_0 = \pi T E_0, \hspace{5mm} \eps_1 = \pi T E_1, \hspace{5mm}
v = \pi T V_0.
\end{equation}
The last equation (\ref{eq:formula}) can be used for the 
practical evaluation
of the transition probability. The remaining task is to calculate
the elements of the monodromy matrices.

\section{Calculation of the transition probability}
\label{sec:calculation}

One may identify several ways to calculate the monodromy matrices 
of the hypergeometric differential equations~\cite{Iwasaki91}. 
Here, we briefly explain the simplest way.

To determine the monodromy matrix, it is crucial to use the integral
representation of the hypergeometric function. By defining
the integral
\begin{equation}
F_{pq}(z) = \int_p^q {\rm d}t \ t^{\alpha-\gamma}(1-t)^{\gamma-\beta-1}
(z-t)^{-\alpha},
\label{eq:integral}
\end{equation}
the following relations hold:
\begin{eqnarray}
& & F_{1\infty} = c_{1\infty} f_0(z;0), \hspace{5mm}
F_{0z} = c_{0z} f_0(z;1-\gamma), \\
& & F_{\infty 0} = c_{\infty 0} f_1(z;0), \hspace{5mm}
F_{1z} = c_{1z} f_1(z;\gamma-\alpha-\beta), \\
& & F_{01} = c_{01} f_{\infty}(z;\alpha), \hspace{5mm}
F_{z\infty} = c_{z\infty} f_{\infty}(z;\beta),
\label{eq:int-inf}
\end{eqnarray}
where $f_0$, $f_1$ and $f_{\infty}$ denote 
the fundamental solutions of the hypergeometric differential equations 
around $z=0, 1, \infty$, respectively.
Here, the constants, $c_{pq}$'s, depend only on
$\alpha$, $\beta$ and $\gamma$, and their explicit expressions
are irrelevant to the present calculation.
By applying the Cauchy's theorem to the integral in eq.~(\ref{eq:integral}),
the following linear relations may be identified:
\begin{eqnarray}
& & F_{01} + F_{1\infty} + F_{\infty 0} = 0, \\
& & F_{01} - F_{0z} + F_{1z} = 0, \\
& & e(\beta-\gamma+1) F_{1\infty} - F_{1z} - e(-\alpha)F_{z\infty} = 0, \\
& & e(\alpha-\gamma)F_{\infty 0} + F_{0z} + F_{z\infty} = 0,
\end{eqnarray}
where $e(\cdot) = \exp(2\pi {\rm i} \cdot)$.
By eliminating $F_{1\infty}$ and $F_{0z}$, we obtain
\begin{eqnarray}
& & (F_{01}, F_{z\infty}) = (F_{\infty 0}, F_{1z}) S, \\
& & S = \frac{1}{e(-\alpha) - e(\beta-\gamma)} \nonumber \\
& & \hspace{5mm} \times
\left( \begin{array}{cc} e(\beta-\gamma) - e(-\gamma) &
e(\alpha+\beta-2\gamma) - e(\beta-\gamma) \\
1-e(-\alpha) & e(\beta-\gamma)-1 \end{array} \right).
\label{eq:Q}
\end{eqnarray}
On the other hand, since the solution pair $(F_{\infty 0}, F_{1z})$
is related to the fundamental solutions around $z=1$, the monodromy
matrix for this pair is easily obtained as
\begin{eqnarray}
\gamma({\rm C}_1) (F_{\infty 0}, F_{1z})
= (F_{\infty 0}, F_{1z}) \Gamma, \hspace{5mm}
\Gamma = \left( \begin{array}{cc} 1 & 0 \\
0 & e(\gamma-\alpha-\beta) \end{array} \right).
\label{eq:Gamma}
\end{eqnarray}
Combining (\ref{eq:Q}) and (\ref{eq:Gamma}), the monodromy
matrix for $(F_{\infty 0}, F_{1z})$ is given as
\begin{eqnarray}
\gamma({\rm C}_1)(F_{01}, F_{z\infty}) = (F_{01},F_{z\infty})
\tilde{R} \\
\tilde{R} = \left(\begin{array}{cc} \tilde{a} & \tilde{b} \\
\tilde{c} & \tilde{d} \end{array} \right) =
S^{-1} \Gamma S.
\end{eqnarray}
This result is easily related to
the fundamental solutions, $f_{\infty}(z;\alpha)$ and
$f_{\infty}(z;\beta)$, by (\ref{eq:int-inf}) as
\begin{equation}
\gamma({\rm C}_1) \left( f_{\infty}(z,\alpha), 
\frac{c_{z\infty}}{c_{01}} f_{\infty}(z,\beta) \right)
= \left( f_{\infty}(z,\alpha), 
\frac{c_{z\infty}}{c_{01}} f_{\infty}(z,\beta) \right) \tilde{R}.
\end{equation}
By comparing the above with the original monodromy matrix 
(\ref{eq:orig-mono}) along
with (\ref{eq:orig-ele}), we finally obtain $a=\tilde{a}$ 
($d=\tilde{d}$). So, as far as $a$ is concerned, we only need to
calculate $\tilde{R}$. This can be performed
by straightforward but slightly lengthy calculation. As a result,
we obtain the matrix element $a$ as
\begin{equation}
a = \frac{e(\beta-\gamma)-e(-\gamma)+e(-\alpha)-1}
{e(\beta-\gamma)-e(\alpha-\gamma)}.
\label{eq:aresult}
\end{equation}
The matrix element $a'$ for the solution $c_2$ is easily obtained
by reversing the sign of $E_0$ and $E_1$ in the result (\ref{eq:aresult}).
From the result for $a$ and $a'$, the transition probability $P$
is obtained from (\ref{eq:formula}) as
\begin{equation}
P = \frac{\sinh^2 (\pi T E_1) \cos^2 (\pi T E_0)}{\sinh^2 (\pi T
\sqrt{E_1^2 + V_0^2})}
+ \frac{\cosh^2 (\pi T E_1) \sin^2 (\pi T E_0)}
{\cosh^2 (\pi T \sqrt{E_1^2 + V_0^2})}.
\label{eq:trans}
\end{equation}

\begin{figure}
\begin{center}
\includegraphics[width=120mm]{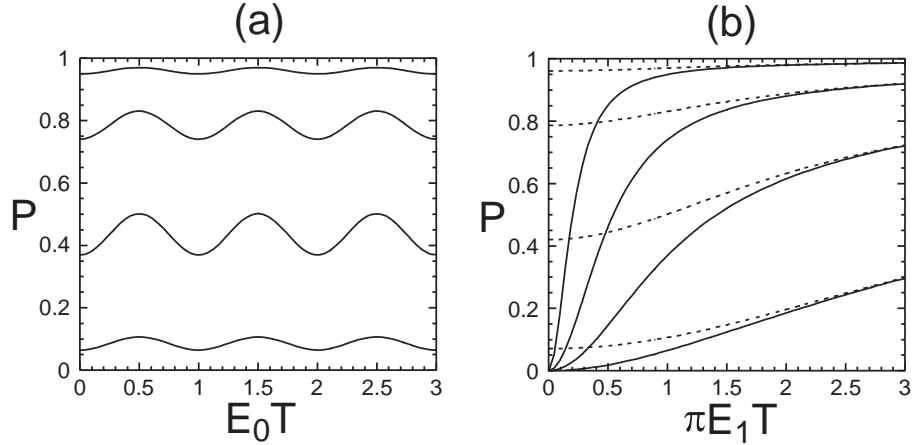}
\end{center}
\caption{\label{fig:graph} (a) The $E_0$-dependence of the transition
probability $P$ for $\pi T E_1 = 1$. 
From the top to the bottom, $\pi V_0 T$ is taken
$0.2, 0.5, 1.0, 2.0$. (b) The $E_1$-dependence of $P$. From the top 
to the bottom, $\pi V_0 T$ is taken $0.2, 0.5, 1.0, 2.0$. The solid(dashed)
lines show the minimum(maximum) value of $P$ at each $E_1$.}
\end{figure}

Let us discuss the nature of this result. The transition probability 
oscillates as the sech-form pulse area, $\pi T E_0$, changes; the transition
probability has minimum and maximum values as a function of $E_0$ as
\begin{equation}
\begin{array}{ll}
P_{\rm min} = \dis{\frac{\sinh^2 (\pi T E_1)}
{\sinh^2 (\pi T \sqrt{E_1^2+V_0^2})}}, & {\rm for} \ \ \pi T E_0 = n\pi, \\
P_{\rm max} = \dis{\frac{\cosh^2 (\pi T E_1)}
{\cosh^2 (\pi T \sqrt{E_1^2+V_0^2})}}, & {\rm for} \ \ \pi T E_0 = (n+1/2)\pi,
\end{array}
\end{equation}
where $n$ is an integer. The oscillation behavior of $P$ for $E_0$ 
is drawn in Fig.~\ref{fig:graph}~(a). 

The amplitude of this oscillation becomes small as $E_1$ increases. 
This feature is shown in Fig.~\ref{fig:graph}~(b).
In the case $E_1 T \gg \max(V_0 T, 1)$, we obtain the ordinary 
Landau-Zener formula
\begin{equation}
P = e^{-\pi V_0^2 T/2E_1}
\end{equation} 
independent of $E_0$.

Finally, we show that the results in the limiting cases 
coincide with the known results.
In the limit $E_1 \rightarrow 0$, the transition probability
is given as
\begin{equation}
P = \frac{\sin^2(\pi T E_0)}{\cosh^2(\pi T V_0)},
\end{equation}
which corresponds to the Rosen-Zener formula~\cite{Rosen32}.
In the limit $E_0 \rightarrow 0$, the transition probability
is given as
\begin{equation}
P = \frac{\sinh^2 (\pi T E_1)}{\sinh^2 (\pi T \sqrt{E_1^2 + V_0^2})}.
\label{eq:formula2}
\end{equation} 
In this case, the present model is related to the second model in
Demkov and Kunike's paper~\cite{Demkov69}, which corresponds to
the form
\begin{eqnarray}
\eps(t) &=& a + b \tanh(t/T), \\
V(t) &=& c.
\end{eqnarray}
Their result for $a=0$ corresponds to the result (\ref{eq:formula2}).

\section{Application of monodromy to multi-level problems}
\label{sec:discussion}

The application of the
monodromy matrix to the transition probability 
is not restricted to the hypergeometric functions.
The monodromy approach is also applicable
to the differential equations whose monodromy is known.
To show such an example, we consider the multi-level problem.
We expect that
more solvable classes can be found by using the present approach.

In this section, we treat the following time-dependent Hamiltonian:
\begin{equation}
H_{ij} = \left\{ \begin{array}{ll} 
\eps(t) & (i=j=1) \\
V_j    & (i=1 \ {\rm and} \ 2\le j \le N) \\
V_i    & (j=1 \ {\rm and} \ 2\le i \le N) \\
0       & ({\rm otherwise})
\end{array} \right. ,
\end{equation}
where the time-dependent part $\eps(t)$ is given as
\begin{equation}
\eps(t) = E_1 \tanh (t/T),
\end{equation}
and $V_j$'s ($2 \le j \le N$) are constants. 
It should be noted that in the limit $E_1 T \rightarrow \infty$, 
this model is reduced to the extended Landau-Zener model 
studied by several authors~\cite{Bohr53,Fano61,Demkov67,Bixon68}.
To eliminate the diagonal element of the Hamiltonian
the wave function denoted by
$\Psi(t) = (a_1, a_2, \cdots, a_N)^{{\rm T}}$ is transformed
into new variables as
\begin{equation}
c_i = \left\{ \begin{array}{ll}
\dis{a_1 \exp \left( {\rm i}\int_0^t \eps {\rm d}t \right)} & (i=1) \\
a_i & (2 \le i \le N)
\end{array} \right. .
\end{equation} 
The integral in the exponent is then calculated as
\begin{equation}
{\rm i}\int_0^t \eps {\rm d}t = {\rm i} E_1 T \log ( \cosh t/T).
\end{equation}
Thus, the Schr\"odinger equation is obtained as
\begin{equation}
T c_{i,t} = \left\{
\begin{array}{ll} 
\dis{\sum_{j=2}^N v_j (\cosh t/T)^{2\eps_1} c_j} & (i=1) \\
v_i (\cosh t/T)^{-2\eps_1} c_1 & (2 \le i \le N)
\end{array} \right. ,
\end{equation}
where
\begin{equation}
\eps_1 = {\rm i}E_1 T/2, \hspace{5mm} v_j = -{\rm i}V_j T.
\end{equation}
By changing the time variable as $z = \sinh(t/T)$,
the equations are modified as
\begin{equation}
\frac{{\rm d}c_i}{{\rm d}z} = \left\{
\begin{array}{ll}
\dis{\sum_{j=2}^{N} v_j (1+z^2)^{\eps_1-1/2} c_j} & (i=1) \\
v_i(1+z^2)^{-\eps_1-1/2} c_1 & (2\le i \le N)
\end{array} \right. .
\end{equation}
We make a further change of variables as
\begin{equation}
d_i = \left\{ \begin{array}{ll}
(1+z^2)^{-\eps_1-1/2}c_1 & (i=1) \\
\dis{\frac{v_i c_i}{z+{\rm i}}
- \lambda_i \left(\eps_1 + \frac12 \right) 
\frac{z-{\rm i}}{z+{\rm i}} d_1}
& (2\le i \le N)
\end{array} \right. ,
\end{equation}
where $\lambda_j$'s are arbitrary constants satisfying
\begin{equation}
\sum_{j=2}^{N} \lambda_j = 1.
\end{equation}
Consequently, we finally obtain
\begin{equation}
(z-{\rm i})\frac{{\rm d}d_1}{{\rm d}z} = - \left(\eps_1+\frac12\right) 
d_1 + \sum_{j=2}^N d_j,
\end{equation}
and for $2\le i \le N$ 
\begin{equation}
(z+{\rm i})\frac{{\rm d}d_i}{{\rm d}z} = \lambda_i
\left(\eps_1^2 + v_i^2 - \frac14 \right) d_1 - d_i 
- \lambda_i \left(\eps_1+\frac12\right) \sum_{j=2}^N d_j.
\end{equation}
This is the Okubo equation expressed by
\begin{equation}
(zI-C) \frac{{\rm d}\vec{d}}{{\rm d}z} = A \vec{d},
\end{equation}
where $I$ is the identity operator, $C$ is a diagonal
matrix, and $A$ is a general matrix. This equation has
been studied by Okubo in detail~\cite{Okubo87}, 
and it is known that 
this form of equation is convenient to study the monodromy.

Thus, it has been shown that at least one specific model of multi-level 
systems can be reduced to the differential equation whose monodromy
is known. Actual calculation of the transition probability needs
explicit treatment of the monodromy matrices, and remains as a 
future problem. The present discussion for multi-level systems 
is preliminary, and more detailed study will be needed to
clarify the efficiency of the monodromy approach.

\section{Summary}
\label{sec:summary}

We have calculated the transition probability for the Hamiltonian
including the tanh-type plus sech-type 
energy difference with constant off-diagonal elements.
The obtained result gives the natural connection between
the known results, the Rosen-Zener model and 
the second Demkov-Kunike model. This model also includes
the Landau-Zener formula in the limit of the large amplitude
of the tanh-type energy difference.

In our calculation, the monodromy of the hypergeometric
functions is essential. We have shown that the monodromy approach 
is also applicable to the multi-level problems. We expect that
the use of the monodromy in the calculation of the transition
probability does not only helps finding more solvable models but 
also connects global properties of the differential equation with
the physical phenomena. Details of calculation especially 
for the multi-level problem remain as future problems.

\ack

We thank H. Nakamura for helpful comments on the model Hamiltonian.
M.L. expresses many thanks to DST(India) and JSPS(Japan) for a JSPS
Invitation Fellowship which enabled him to visit
Osaka City University during October-November 2002.
His work is also supported by the Department of Science and
Technology(DST), Government of India.
K.N. and T.K. are also grateful to JSPS for the financial support of
the Fundamental Research, C-2, No. 13640391, entitled "Quantum Transport
in Quantum Chaotic Systems."

\appendix

\section{Solvable classes}
\label{app:class}

The model considered in the main part of this paper belongs
to one solvable class called class 1 below.
It can be given as
\begin{eqnarray}
\eps(t) &=& \frac{E_0 T + E_1T y}{1+y^2} \frac{{\rm d}y}{{\rm d}t}, \\ 
V(t) &=& \frac{V_0 T}{\sqrt{1+y^2}} \frac{{\rm d}y}{{\rm d}t},
\end{eqnarray}
where $y(t)$ is an `arbitrary' monotonically increasing function
satisfying $y(t) \rightarrow \pm \infty$ for $t \rightarrow \pm \infty$.
When we adopt $y(t) = \sinh(t/T)$, we obtain (\ref{eq:eps}) and 
(\ref{eq:v}). For this class, the Schr\"odinger equation can be 
reduced to the same hypergeometric differential equation (\ref{eq:Gauss}) 
through the change of variable 
$z(t) = (y(t)+{\rm i})/2{\rm i}$~\cite{Ishkhanyan00}.
Hence, all models of this class give the same transition probability 
(\ref{eq:trans}). In this class, however, we have to
define the transition probability carefully. 
In the limit $t \rightarrow -\infty$
($y \rightarrow -\infty$), the matrix elements become
\begin{eqnarray}
\eps(t) &\rightarrow& \frac{E_1 T}{y} \frac{{\rm d}y}{{\rm d}t}, \\
V(t) &\rightarrow& - \frac{V_0 T}{y} \frac{{\rm d}y}{{\rm d}t}.
\end{eqnarray}
Hence, the wave function of the ground state 
in this limit has mixed components 
as treated in \S~\ref{sec:monodromy}. The initial state is
taken as the ground state in this limiting Hamiltonian, and
the transition probability is defined as square of modulus
of the final amplitude of the excited states.

The application of the monodromy is not restricted to
the class 1. As discussed by Ishkhanyan~\cite{Ishkhanyan00},
as long as the complex path $z(t)=(y(t)+{\rm i})/2{\rm i}$ is used,
the calculation by the monodromy is efficient. For example, the
following solvable class can be considered:
\begin{eqnarray}
\eps(t) &=& \frac{E_0 T + E_1 T y}{1+y^2} \frac{{\rm d}y}{{\rm d}t}, \\
V(t) &=& \frac{V_0 T}{1+y^2} \frac{{\rm d}y}{{\rm d}t}.
\end{eqnarray}
This class, called here the class 2, has been first studied 
by Carroll and Hioe~\cite{Carroll86}.
There, the transition probability has been calculated by solving
the Riemann-Papperitz equation without resorting to the monodromy.
By following the Ishkhanyan's discussion, however, 
our monodromy approach is also efficient for the class 2,
and gives an alternative method.

\end{document}